# Anomalous frequency-dependent ionic conductivity of lesion-laden human-brain tissue

David Emin,[1] Massoud Akhtari,[2] Aria Fallah,[3] Harry V. Vinters,[4] and Gary W. Mathern[5]

[1] Department of Physics and Astronomy, University of New Mexico, Albuquerque, NM 87131, USA

[2] Semel Institutes for Neuroscience and Human Behavior, David Geffen School of Medicine, University of California at Los Angeles, Los Angeles, CA 90096, USA

[3] Departments of Neurosurgery and Pediatrics, David Geffen School of Medicine, University of California at Los Angeles, Los Angeles, CA 90096, USA

[4] Department of Pathology, David Geffen School of Medicine, University of California at Los Angeles, Los Angeles, CA 90096, USA

[5] Departments of Neurosurgery and Psychiatry and Biobehavioral Medicine, David Geffen School of Medicine, University of California at Los Angeles, Los Angeles, CA 90096, USA

We study the effect of lesions on our four-electrode measurements of the ionic conductivity of (~1 $cm^3$) samples of human brain excised from patients undergoing pediatric epilepsy surgery. For most (~94%) samples the low-frequency ionic conductivity rises upon increasing the applied frequency. We attributed this behavior to the long-range (~0.4 mm) diffusion of solvated sodium cations before encountering impenetrable blockages such as cell membranes, blood vessels and cell walls. By contrast, the low-frequency ionic conductivity of some (~6 %) brain-tissue samples falls with increasing applied frequency. We attribute this unusual frequency-dependence to the electric-field induced liberation of sodium cations from traps introduced by the unusually severe pathology observed in samples from these patients. Thus, the anomalous frequency-dependence of the ionic conductivity indicates trap-producing brain lesions.

## I. INTRODUCTION

We address the frequency-dependent modulation of the room-temperature electrical conductivity of (~1 $cm^3$) samples of human-brain tissue that are freshly excised from 67 pediatric patients undergoing epilepsy surgery.[1,2] In most instances the low electrical conductivity increases slowly as the frequency of the applied electric field is increased (6 – 1000 Hz). We ascribed this electrical conduction to moderate densities ($2.5 \times 10^{25}$ $m^{-3}$) of solvated sodium cations which diffuse with very low mobility (~ $4 \times 10^{-8}$ $m^2$/V-s) under the influence of an applied electric field until they encounter an impenetrable blockage.[1] The separation between such barriers is found to be about 0.4 millimeters. In these cases the small diffusion constants of solvated sodium cations (~ $10^{-9}$ $m^2$/s) are nearly invariant.[2] Thus differences in the magnitudes of the conductivities of these samples result primarily from differences in their sodium cation concentrations.[2]

Nonetheless, we find a qualitatively different frequency dependence in samples from four patients. In these instances the electrical conductivity *decreases*, rather than increases, as the frequency of the applied electric field is increased. Plots *a* and *b* of Fig. 1 show two examples of these anomalous frequency dependences. For comparison plot *c* of Fig. 1 shows an example of the behavior we usually observe.

Rather than coming from a general human or animal population, our samples are freshly excised from patients undergoing epilepsy surgery. Furthermore, samples that manifest the “anomalous” frequency-dependence of their low-frequency conductivities are a small subset of our restricted population. This subset was acquired from over a decade of conductivity measurements.

Samples excised from these patients indicated atypically severe damage. As illustrated in Fig. 2, these samples showed substantial regions devoid of gray or white matter cytology and cytroarchitecture. In particular, histopathological examinations of brain tissue from three of the four patients whose conductivities were anomalous showed severe brain injury due to large territory perinatal infarcts with surrounding gliosis. The fourth patient had moderate focal cortical dysplasia with areas of severe cytoarchitectural abnormality where the lamination and organization of cortical neurons was disrupted.

We now explain how human brain tissue containing disease-related injury/abnormality can produce the unusual frequency-dependence of sodium cations' conductivity. To begin, as schematically illustrated in Fig. 3, sodium cations tend to accumulate in lesion-laden regions. Indeed, sodium-MRI imaging shows that the sodium concentrations are increased by ~ 40% and ~ 25% within and adjacent to a tumor, respectively.[7-9] The application of an electric field then extricates some sodium cations from these traps. Indeed, the electric-field induced freeing of electronic charge carriers from shallow traps is a well-known phenomenon in insulators and semiconductors.[10] We now show that the liberation of sodium cations from their traps, and hence their ionic conductivity, decreases as the frequency of the applied electric field is increased.

## II. CALCULATION

The electrical conductivity is proportional to the density of mobile solvated sodium cations. Thus the time dependence of the transient conductivity $\sigma_t(t)$ of sodium cations is just that for their being free to move following the application of an electric field at $t = 0$, $P_f(t)$. The application of an electric field enhances sodium cations' (1) liberation from traps thereby enabling cation movement and (2) flow to blockages which prevent further cation movement.

The master equations governing these processes are:

$$\frac{dP_f(t)}{dt} = R_l P_t - R_b P_f, \quad (1)$$

$$\frac{dP_t(t)}{dt} = -R_l P_t, \quad (2)$$

and

$$\frac{dP_b(t)}{dt} = R_b P_f, \quad (3)$$

where $P_f(t)$, $P_t(t)$ and $P_b(t)$ represent the probabilities of sodium cations being free, trapped and blocked at time $t$. The rates for electric-field-induced liberation of cations from traps and free cations being blocked are denoted by $R_l$ and $R_b$, respectively.

The requirement that probability be conserved, $P_f$, + $P_t$ + $P_b$ = 1, is now utilized to eliminate $P_t(t)$ from the first of the master equations:

$$\frac{\partial P_f(t)}{\partial t} = R_l\left(1 - P_f - P_b\right) - R_b P_f = R_l(1 - P_b) - (R_l + R_b)P_{f.} \quad (4)$$

A single second-order differential equation of just one dependent variable $P_f$ is obtained by differentiating this equation with respect to the time *t* and then employing the third of the original master equations to eliminate the variable $P_b$:

$$\frac{d^2P_f(t)}{dt^2} + (R_b + R_l + R_c)\frac{dP_f(t)}{dt} + (R_l R_b)P_f = 0. \quad (5)$$

The general solution of this differential equation is[11]

$$P_f(t) = Ae^{\lambda_+ t} + Be^{\lambda_- t}, \quad (6)$$

where *A* and *B* are constants that are determined by our problem's initial conditions and

$$\lambda_\pm \to \frac{-(R_b + R_l) \pm \sqrt{(R_b - R_l)^2}}{2}. \quad (7)$$

Then the two solutions simply become $\lambda_+ = -R_l$ and $\lambda_- = -R_b$. The constants *A* and *B* are determined from the initial conditions.

The probability of carriers initially being free is

$$P_f(0) \equiv A + B = 1 - f, \quad (8)$$

where *f* denotes the fraction of carriers initially trapped at inclusions. In addition, the initial rate of change of the probability of carriers being free is:

$$\dot{P}_f(0) = -AR_l - BR_b = fR_l \ - (1-f)R_b\,. \quad (9)$$

The simultaneous equations, Eq. (8) and Eq. (9), are readily solved to yield the constants:

$$A = \frac{-fR_l}{R_l - R_b} \quad (10)$$

and

$$B = \frac{R_l - (1-f)R_b}{R_l - R_b}. \quad (11)$$

Incorporating these values of *A* and *B* into Eq. (6) and inserting its two roots yields

$$P_f(t) = \frac{[R_l - (1-f)R_b]e^{-R_b t} - fR_l e^{-R_l t}}{R_l - R_b}. \quad (12)$$

In the absence of sources for electric-field-induced liberation of ionic carrier, $f = 0$, the conductivity $\sigma_t(t)$ falls monotonically with time since then $P_f(t) = \exp(-R_b t)$. A different situation can prevail in the presence of sources of liberated ionic charge carriers, $f \neq 0$. For example, in the extreme situation for which $f = 1$:

$$P_1(t) = \frac{R_l}{R_l - R_b}\left(e^{-R_b t} - e^{-R_l t}\right). \quad (13)$$

At sufficiently short times $P_1(t)$ rises with time as carriers are liberated from traps: $P_1(t) \rightarrow R_l t$ as $t \rightarrow 0$. However, at sufficiently long time $P_1(t)$ falls with time as liberated carriers are progressively stopped by blockages. For example, when this transient relaxation is limited by cations' electric-field-induced diffusion to barriers, $R_l >> R_b$, then $P_1(t) \rightarrow \exp(-R_b t)$ as $t \rightarrow \infty$. Similarly, when this transient relaxation is limited by cations' electric-field-induced extraction from traps, $R_l << R_b$, then $P_1(t) \rightarrow (R_l/R_b) \exp(-R_l t)$ as $t \rightarrow \infty$.

Figure 4 shows plots of the calculated $P_f(t)$ from Eq. (12) versus $R_b t$ with $R_l/R_b = 10$ for $f = 0$, 0.25, 0.5, 0.75 and 1. This family of curves illustrates how altering $f$, the fraction of ions that are trapped at $t = 0$, affects the probability of their being free at later times. Without trapping, $f = 0$, $P_f(t)$ falls monotonically with $t$ as ions progressively encounter blockages. By contrast, electric-field-induced extraction of ions from traps causes $P_f(t)$ to increase with time until a significant fraction of ions are freed. Ultimately, at extremely long times, $P_f(t)$ decreases with time with a drastically reduced $f$-dependence as the freed ions diffuse until they are stopped at blockages.

The frequency-dependent conductivity $\sigma(\omega)$ is directly obtained from the time-dependence of the transient ionic conductivity $\sigma_t(t)$ that follows the imposition of an electric field at $t = 0$:[1]

$$\sigma(\omega) = \omega \int_0^{\infty} dt \sigma_t(t) sin(\omega t). \quad (14)$$

Recalling that the time-dependence of $\sigma_t(t)$ is the same of that of $P_f(t)$, Eq. (12) is inserted into Eq. (14) and evaluated using a standard integral to yield[12]

$$\sigma(\omega) = \frac{\sigma_0}{(1-f)} \left\{ \left[ \frac{R_l - (1-f)R_b}{R_l - R_b} \right] \left[ \frac{1}{1 + (R_b/\omega)^2} \right] - \left( \frac{f R_l}{R_l - R_b} \right) \left[ \frac{1}{1 + (R_l/\omega)^2} \right] \right\}, \quad (15)$$

where $\sigma_0$ is the conductivity without either carrier-liberating traps or carrier-stopping blockages.

Figure 5 shows plots of $\sigma(\omega)[(1 - f)/\sigma_0]$ as calculated from Eq. (15) versus $\omega/R_b$ with $R_l/R_b = 10$ for $f = 0$, 0.25, 0.5, 0.75 and 1. At extremely low frequencies, well below those of our measurements, $\sigma(\omega)$ manifests almost no $f$-dependence as transport is limited by ions encountering blockages. The $f$-dependence as $\sigma(\omega)$ emerges at higher frequencies where $\sigma(\omega)$ is governed by the electric-field-induced extraction of ions from traps. The conductivity then falls ever more steeply with increasing frequency $\omega$ as $f$, the fraction of ions that are initially trapped, is increased. Distinctively, when an insignificant fraction of ions are initially trapped, $f = 0$, $\sigma(\omega)$ increases with the applied-frequency $\omega$.

The range of the abscissa and the ratio $R_l/R_b$ utilized in Figs. 4 and 5 were chosen to illustrate the transitions between regimes with weak and strong dependencies on the fraction of trapped cations $f$. The value of $R_b$ estimated from the data of Ref. (1) and the minimum value of our measurement frequency 6 Hz imply that the data of Fig. 1 correspond to $\omega/R_b >> 1$. In this circumstance the transition from positive to negative values of $\partial\sigma(\omega)/\partial\omega$ occurs for very small values of $f$. Alternatively stated, trapping of only a small fraction of mobile sodium cations produces the transition from the normal to the “anomalous” frequency dependence of the ionic conductivity. Thus, the frequency dependence of the conductivity is a very sensitive probe of the presence of ionic traps which we attribute to severely damaged human brain tissue.

## III. DISCUSSION

Typically the electrical conductivities of our samples of excised brain tissue fall with increasing time and decreasing applied frequency.[1,2] This behavior was attributed to solvated sodium cations freely diffusing long distances until they encounter impenetrable blockages. The conductivity rises as the frequency with which the direction of the applied electric field reverses is raised because increasing carriers’ back-and-forth movement in response to the alternating electric field forestalls their being stopped at blockages.

By contrast, in a few instances the ionic conductivity is observed to fall with increasing frequency. This unusual behavior is attributed to ions being trapped within lesion-laden tissue. As the frequency with which the direction of the applied electric-field alternates is increased there is a decreasing probability of the electric field extracting ions from these traps.

For clarity, we have employed the simplest of mathematical models even though human brain tissue is inhomogeneous.[7,8] The inhomogeneity can be crudely modeled by presuming a distribution of capture and release rates for traps. Indeed, we previously modeled the free diffusion of sodium cations between blockages in typical excised human-brain tissue with a distribution of diffusion lengths.[1] Nonetheless, the essential distinction between (1) blockage-limited sodium cation diffusion and (2) electric-field-induced freeing of sodium cations from traps survives such averaging. In particular, sodium cation diffusion between blockages yields a frequency-dependent conductivity that increases with increasing applied frequency. By contrast, electric-field-induced freeing of these ions from traps yields a frequency-dependent conductivity that decreases with increasing applied frequency.

As the examples in Fig. 1 show, the net variation among all samples' room-temperature ionic conductivities is about a factor of two.[2] As illustrated in Fig. 3, most sodium cations are free to diffuse. Only a small fraction of ions are blocked by barriers or trapped. Nonetheless, the frequency-dependence of the conductivity is sensitive to changes in the densities of these subsets of ions. The blockage of ions and their extrication from lesion-laden regions produce opposite frequency dependences. Indeed, the frequency-dependent conductivities of some of our samples measured between 6 and 1000 Hz even manifest the non-monotonic frequency dependence expected of these two effects acting in tandem.

In summary, our measurements and analyses of excised samples of human brain address how abnormalities in brain tissues affect its ionic transport. For (~1 $cm^3$) samples of very severely compromised brain tissue the measured ionic conductivity manifests an atypical dependence on the frequency of the applied electric field. In particular, this conductivity *decreases* with increasing frequency. Here we have shown that the electric-field-induced extraction of sodium cations from traps produces this frequency dependence. Thus, we explain the decrease of brain tissues' ionic conductivity with increasing frequency as indicating lesion-related sodium-cation traps.

## ACKNOWLEDGEMENTS

This work was supported by the Weil Fund at UCLA Semel Institutes for Neuroscience and Human Behavior and NIH R21 NS060675-01 and NIH RO1 NS38992.

## Figure Captions

Fig. 1. In sub-figures *a* and *b* the frequency-dependent conductivities (in S/m) of samples excised from two patients with severe brain injury and/or abnormalities are plotted against the applied frequency (in Hz). Distinctively, these conductivities fall as the applied frequency is increased. For comparison, in sub-figure *c* the frequency-dependent conductivity (in S/m) of a sample excised from a typical patient is plotted versus frequency (in Hz). Here the conductivity rises slowly as the applied frequency is increased. Detailed descriptions of our four-probe conductivity measurements on freshly excised cranial and brain tissue are found in Refs. 3-6.

Fig. 2. A micrograph of brain tissue from a patient whose conductivity manifests the anomalous frequency dependence (in pink stain on the left) is compared with a micrograph with equal magnification from one of our typical pediatric epilepsy patients (in blue stain on the right). The micrograph on the left is highly disorganized. Its small dark dots characteristic of grey matter are relatively sparse and inhomogeneous. The fine meshwork characteristic of white matter is even absent in the lower portion of its image.

Fig. 3 Subfigure *a* schematically illustrates ions that are free to diffuse (solid dots) between widely separated blockages (solid arcs). Subfigure *b* depicts the buildup of ions at or near a brain lesion (darkened circular region). The application of an electric field extricates ions from the abnormal region thereby augmenting the ionic conductivity.

Fig. 4. The transient probability of ions moving freely after an electric field is applied at $t = 0$ given by Eq. (12) $P_f(t)$ is plotted versus $R_b t$ with $R_l/R_b = 10$ for $f = 0, 0.25, 0.5, 0.75$ and 1. In the absence of initial trapping, $f = 0$, $P_f(t)$ falls monotonically as ions progressively encounter

blockages at the rate $R_b$. By contrast, when a significant fraction of ions are initially trapped, $P_f(t)$ first rises as ions are liberated from their traps and then falls as liberated ions are stopped by blockages. The circled regime indicates that we associate our measurements with very small values of $R_b t$.

Fig. 5. The normalized frequency-dependent conductivity of Eq. (15) $\sigma(\omega)[(1-f)/\sigma_0]$ is plotted versus $\omega/R_b$ with $R_l/R_b = 10$ for $f = 0$, 0.25, 0.5, 0.75 and 1. In the absence of initial trapping, $f = 0$, $\sigma(\omega)$ rises monotonically with increasing $\omega$ as freed ions progressively encounter blockages at the rate $R_b$. Similarly, $\sigma(\omega)$ also rises with increasing $\omega$ at low enough frequencies when a significant fraction of ions is initially trapped, $f \neq 0$. The concomitant negligible dispersion indicates that this very low frequency ionic transport is being limited by blockages. However, the extraction of ions from their traps manifests itself at higher frequencies where $\sigma(\omega)$ falls as $\omega$ increases. The concomitant strong $f$-dependence indicates that transport in this domain is primarily determined by liberations of ions from their traps. The circled regime indicates that we associate our measurements with very large values of $\omega/R_b$.

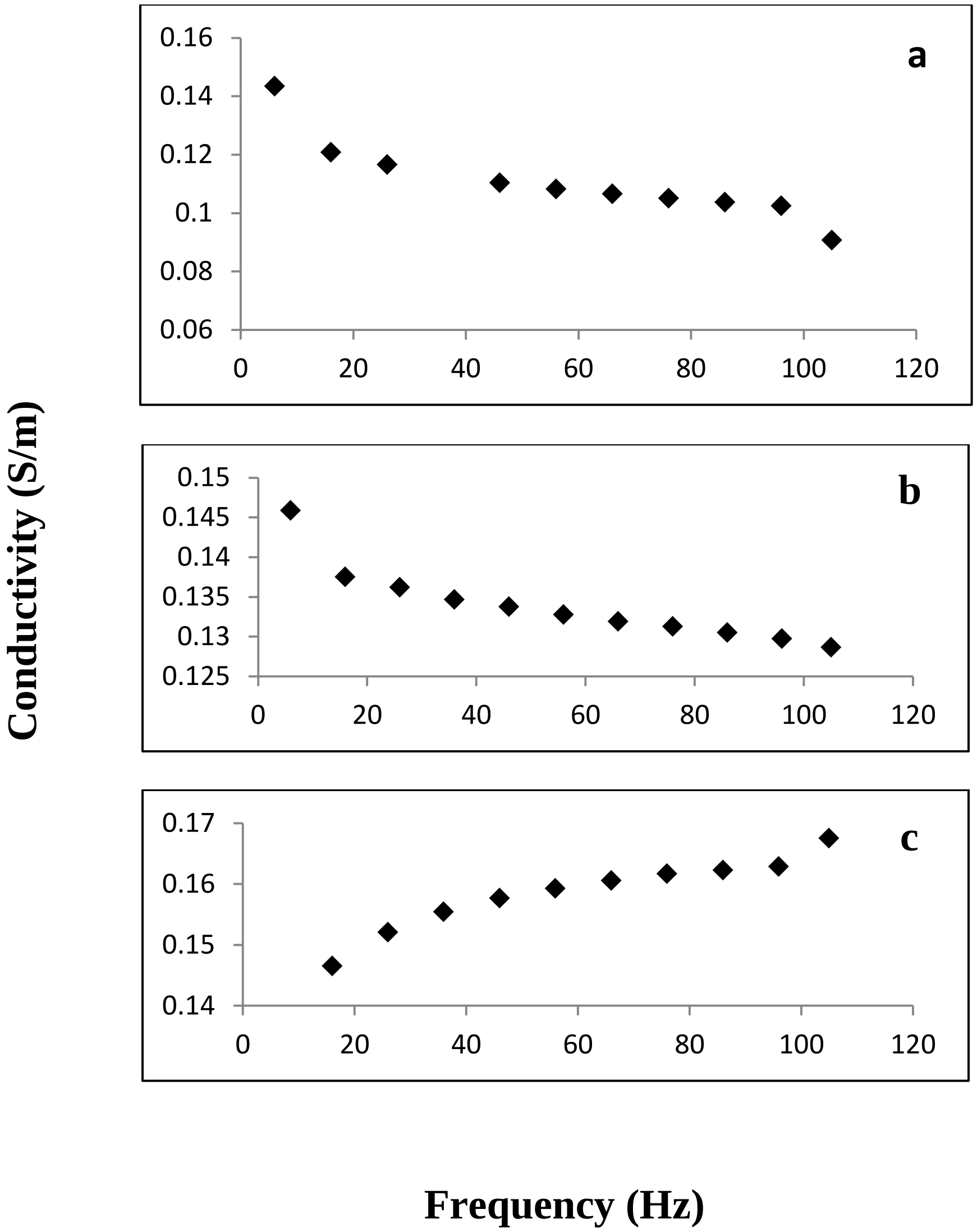

Figure 1

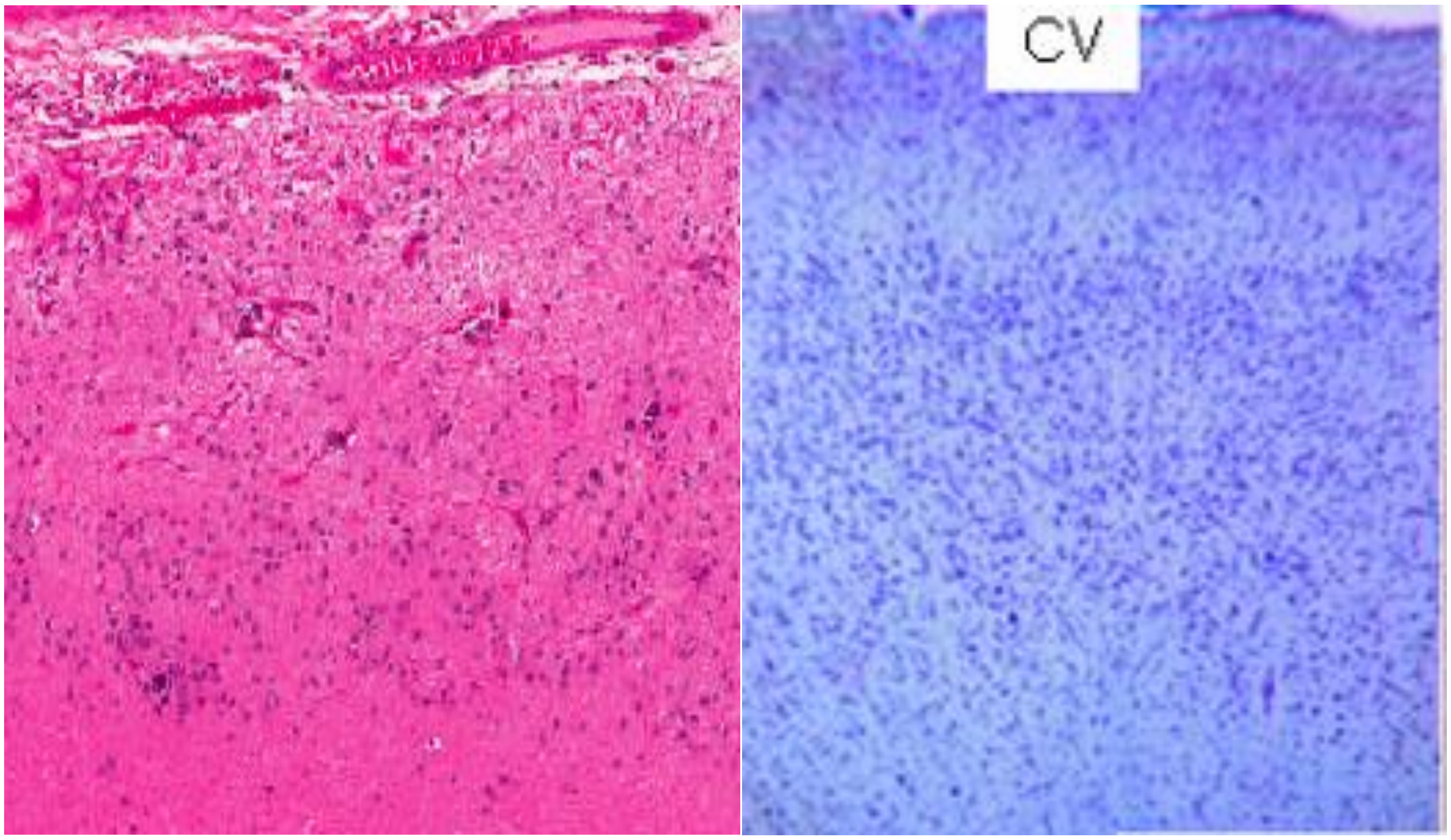

Figure 2

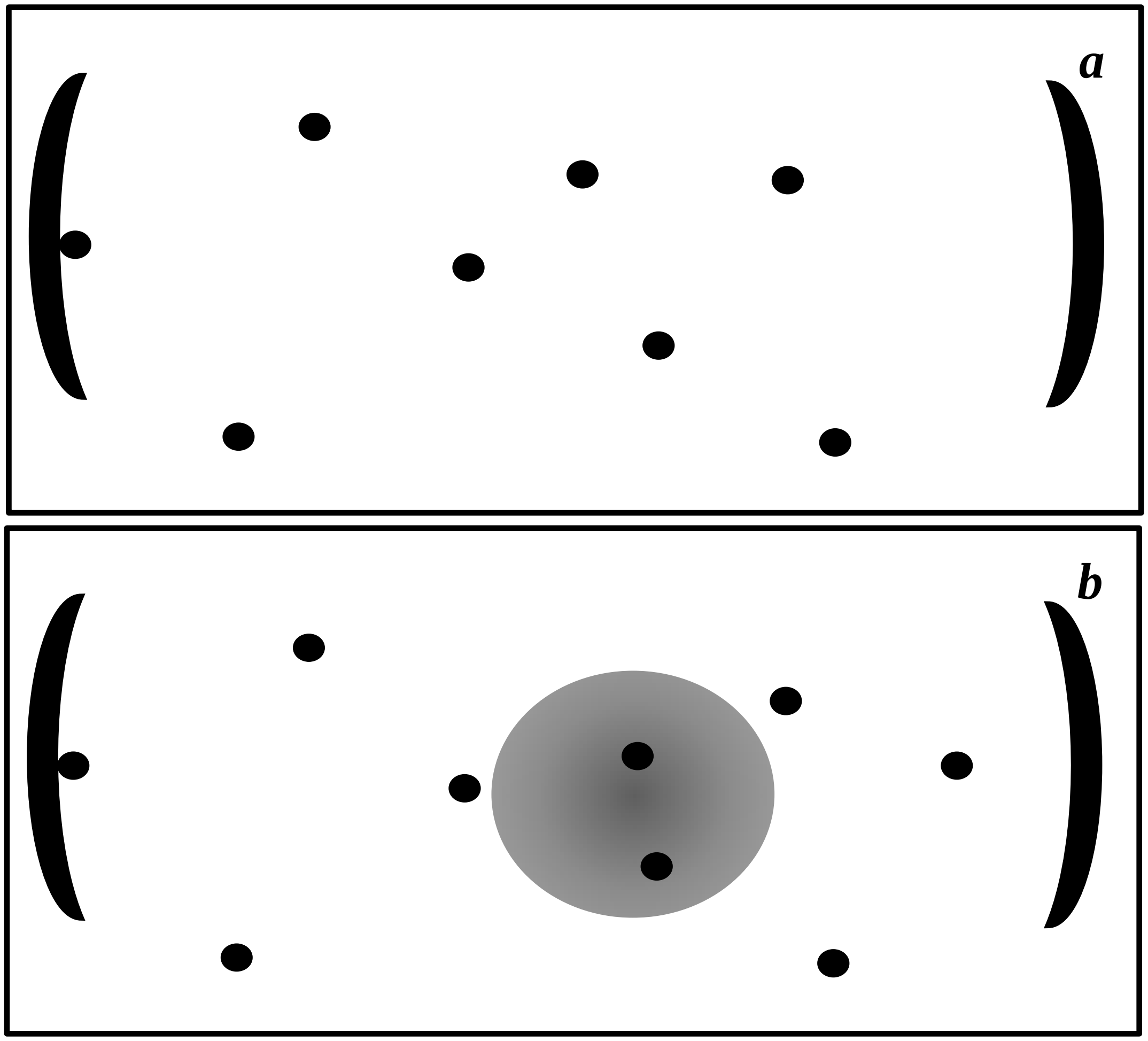

Figure 3

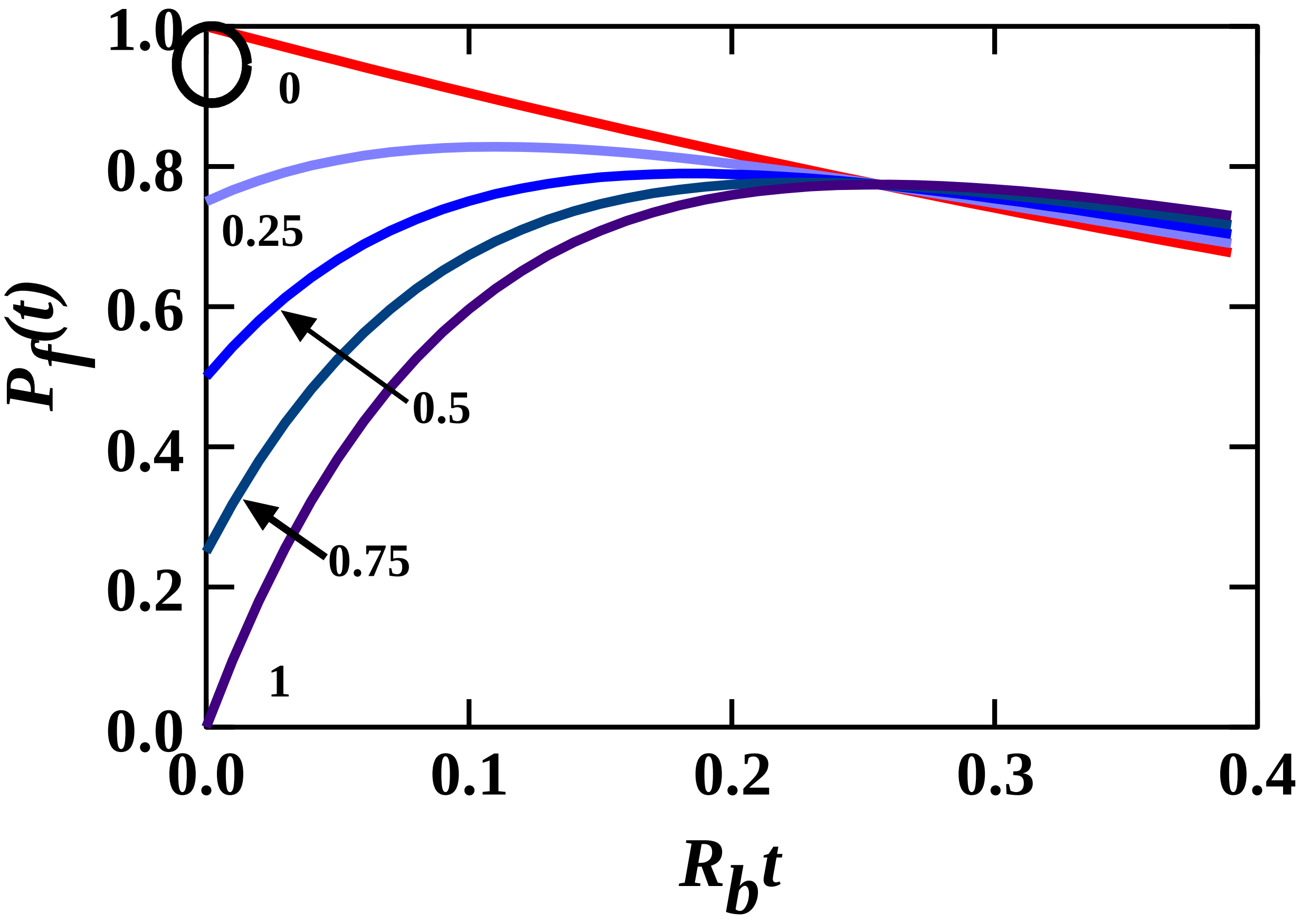

Figure 4

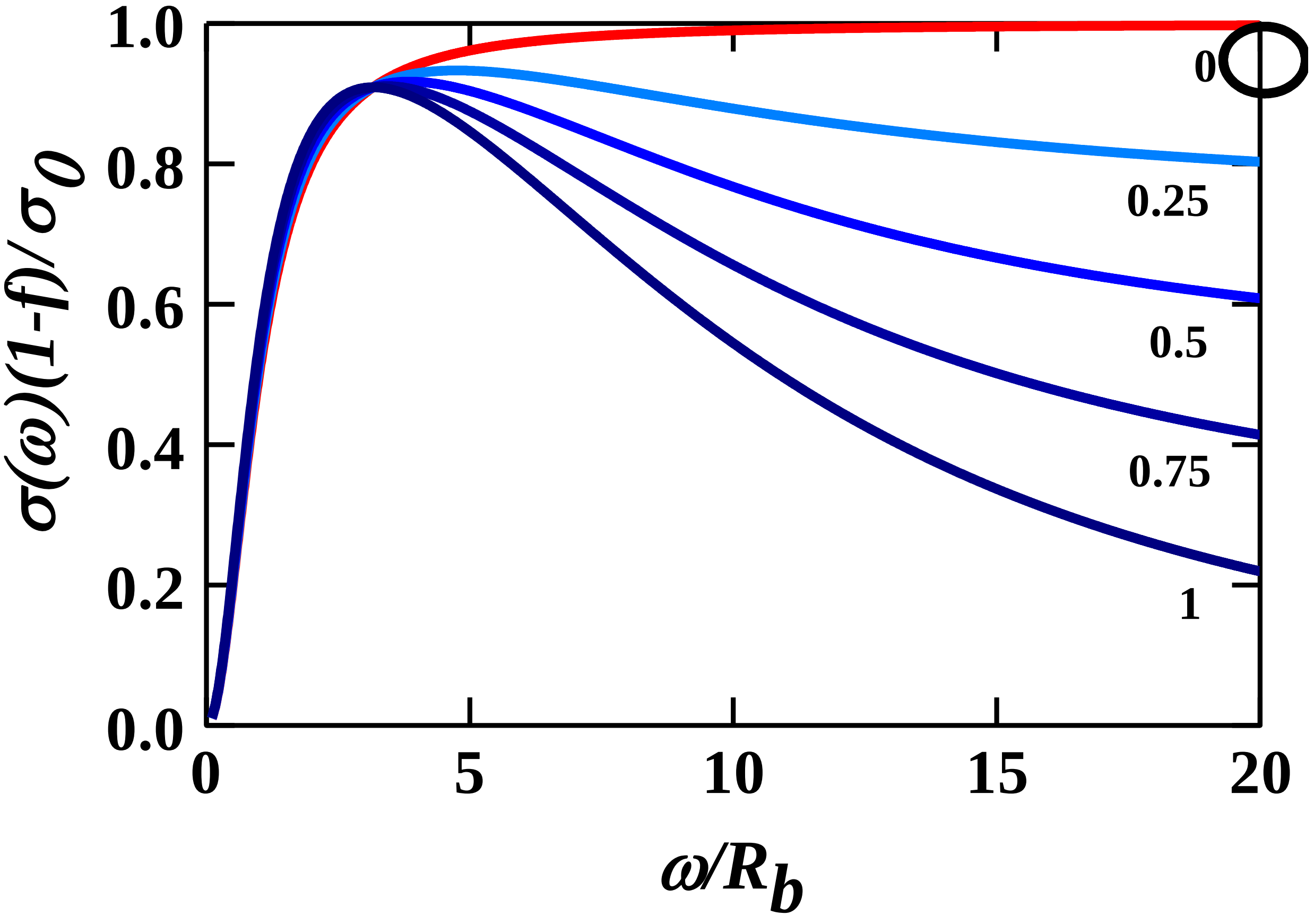

Figure 5